\documentclass[letterpaper, 10pt, conference]{ieeeconf}  
\IEEEoverridecommandlockouts 
\usepackage{cite}
\usepackage{amsmath,amssymb,amsfonts}
\usepackage{algpseudocode}
\usepackage{graphicx}
\usepackage{textcomp}
\usepackage{xcolor}
\def\BibTeX{{\rm B\kern-.05em{\sc i\kern-.025em b}\kern-.08em
    T\kern-.1667em\lower.7ex\hbox{E}\kern-.125emX}}
    
\usepackage[utf8]{inputenc}
\usepackage{csquotes}
\usepackage{amsmath}
\usepackage{mathtools}
\usepackage{amsfonts}
\usepackage{amssymb}

\usepackage{amsthm}
\usepackage{etoolbox}
\usepackage{siunitx}
\usepackage[hyphens]{url}
\usepackage[hidelinks, bookmarks=false]{hyperref}

\title{\LARGE \bf Proportional Neuro-Impulsive Control of SISO LTI Systems}

\renewcommand{\authorrefmark}[1]{\raisebox{0pt}[0pt][0pt]{\textsuperscript{\footnotesize #1}}}
\newcommand{\authorrefmarktanks}[1]{\raisebox{0pt}[0pt][0pt]{\textsuperscript{#1}}}
\author{Christian Fernandez Lorden\authorrefmark{1}\and Elena Petri\authorrefmark{2}\and W.P.M.H. (Maurice) Heemels\authorrefmark{2}\and Alessio Franci\authorrefmark{1,3}%
\thanks{\authorrefmarktanks{1}C.\ Fernandez Lorden and A.\ Franci are with the Department of Electrical Engineering and Computer Science, University of Liège, 4000 Liège, Belgium. \authorrefmarktanks{2}E.\ Petri and W.P.M.H. Heemels are with the Department of Mechanical Engineering, Eindhoven University of Technology, 5612 Eindhoven, the Netherlands. \authorrefmarktanks{3}A.\ Franci is with the WEL-T Department, WEL Research Institute, 1300 Wavre, Belgium. \{christian.fernandezlorden, afranci\}@uliege.be, \{e.petri, W.P.M.H.Heemels\}@tue.nl.}
}
\date{May 2026}

\newcommand{\R}{\mathbb{R}}
\newcommand{\N}{\mathbb{N}}
\DeclareFontFamily{U}{matha}{\hyphenchar\font45}
\DeclareFontShape{U}{matha}{m}{n}{
<-6> matha5 <6-7> matha6 <7-8> matha7
<8-9> matha8 <9-10> matha9
<10-12> matha10 <12-> matha12
}{}
\DeclareSymbolFont{matha}{U}{matha}{m}{n}

\DeclareFontFamily{U}{mathx}{\hyphenchar\font45}
\DeclareFontShape{U}{mathx}{m}{n}{
<-6> mathx5 <6-7> mathx6 <7-8> mathx7
<8-9> mathx8 <9-10> mathx9
<10-12> mathx10 <12-> mathx12
}{}
\DeclareSymbolFont{mathx}{U}{mathx}{m}{n}
\DeclareMathDelimiter{\vvvert} {0}{matha}{"7E}{mathx}{"17}%

\newcommand\swapifbranches[3]{#1{#3}{#2}}
\makeatletter
\MHInternalSyntaxOn
\patchcmd{\DeclarePairedDelimiter}{\@ifstar}{\swapifbranches\@ifstar}{}{}
\MHInternalSyntaxOff
\makeatother

\DeclarePairedDelimiter{\vnorm}{\lVert}{\rVert}
\DeclarePairedDelimiter{\mnorm}{\vvvert}{\vvvert}

\DeclareMathOperator{\dom}{dom}

\newcommand{\dwellt}{{{\mathrm{T}}}} 
\newcommand{\impactmap}{\Phi}              
\newcommand{\dop}[1]{\mathop{\mathrm{d}}#1} 
\newcommand{\eqdefsymbol}{\coloneqq}

\newtheorem{theorem}{Theorem}
\newtheorem{lemma}[theorem]{Lemma}

\newtheorem{assumption}[theorem]{Assumption}
\newtheorem{definition}[theorem]{Definition}

\newtheorem{remark}[theorem]{Remark}

\newif\ifwithappendix
\withappendixtrue

\makeatletter
\g@addto@macro\@floatboxreset{\vspace*{5pt}}
\patchcmd{\table}{\@setminipage}{\@setminipage\vspace*{5pt}}{}{\PackageWarning{main}{table patch failed}}
\expandafter\patchcmd\csname table*\endcsname{\@setminipage}{\@setminipage\vspace*{5pt}}{}{\PackageWarning{main}{table* patch failed}}
\patchcmd{\@maketitle}{\vskip0.25in}{\vskip0.25in\vskip3pt}{}{\PackageWarning{main}{title patch failed}}
\makeatother

\begin{document}
\bstctlcite{BSTcontrol} 

\maketitle

\begin{abstract}
    This paper explores a controller that is at the crossroads of neuromorphic, i.e., spike-based, and impulsive control systems.
    The controller consists of a single integrate-and-fire-like neuron that generates an input proportional to the system output when the neuron variable reaches a threshold.
    We prove that any static output-feedback stabilizable single-input single-output (SISO) linear time-invariant (LTI) plant can be made semi-globally asymptotically stable (sGAS) in closed loop with the proposed neuro-impulsive controller.
    This improves on existing fixed-amplitude neuromorphic controllers, which only achieve practical stability, but at the price of using graded spiking inputs.
    Moreover, we provide an algorithm for choosing appropriate controller parameters given a set of admissible initial conditions.
    A numerical example of an inverted pendulum linearized around its upright equilibrium illustrates the method.
\end{abstract}

\section{Introduction}\label{sec:introduction}

Understanding and reproducing biological control laws and feedback loops is an old but ever-reappearing goal in control and in science at large.
Norbert Wiener's cybernetics \cite{wiener_cybernetics_2019} is one well-known attempt, trying to unify the study of feedback loops, whether biological, physical, or even social.
In a similar spirit, control engineers have recently taken an interest in neuromorphic engineering, a field that studies how to reproduce neuronal dynamics in physical substrates and how to use them to compute, sense, and actuate \cite{mead_neuromorphic_1990}.
The field started as an electronics subfield and has achieved its most impressive results there, but its link with control has been clear since its inception~\cite{deweerth_simpleneuronservo_1991} and has recently attracted new interest \cite{ribar_neuromorphiccontrol_2021, medvedeva_formalizing_2025, petri_emulationneuromorphic_2025, eilers_eventbasedstability_2026}.

Neuron models come in many shapes and forms \cite{izhikevich_modelszoo_2004}, and so do neuromorphic controllers.
Most rely on a common idea: a dynamic element, the neuron, evolves according to its input until a threshold is reached; a discrete event, the spike, is then triggered and the element is reset.
The spike may be a true discrete event produced by a hard threshold, as in the integrate-and-fire models \cite{burkitt_ifreview_2006}, or embedded in the dynamics as a fast but continuous trace, as in conductance-based and mixed-feedback models \cite{franci_balanceequation_2013, sepulchre_controlacrossscales_2019, ribar_neuromorphiccontrol_2021}.
The models differ greatly, but they all agree on one point: the neuron produces a control pulse of fixed amplitude (almost fixed for continuous spikes) at each spike.

Controlling with discrete events is not unique to neuromorphic engineering.
As early as 1963, integral pulse frequency modulation (IPFM) \cite{li_IPFM_1963} described controllers that, also inspired by neural communication, generate fixed-amplitude pulses with a frequency proportional to the input signal, and that line of work went as far as characterizing the resulting practical stability zone \cite{king-smith_IPFM-ultimatebound_1969}.
This is close, in spirit, to controllers relying on an integrate-and-fire neuron as a spike generator \cite{petri_lifhybridanalysis_2024, petri_emulationneuromorphic_2025, eilers_eventbasedstability_2026}, as both trigger an event when the integral of an error reaches a threshold.
In a similar way, impulsive control was created to analyze systems that follow continuous dynamics most of the time but undergo discrete transitions at time instants that can either be fixed or depend on the state \cite{haddad_impulsive_2006}.

The way neuromorphic controllers have largely been conceptualized leaves at least two open problems.
Since every spike has the same amplitude, the correction sent to the plant does not scale with the plant's distance from its target.
Close to the desired unstable fixed point, fixed-amplitude spikes can therefore only lead to practical stability, with oscillation being unavoidable \cite{petri_emulationneuromorphic_2025, eilers_eventbasedstability_2026}.
Far from it, spikes must be delivered many times to have a large effect, so the spike frequency grows with the distance to the target and the inter-spike times can only be bounded semi-globally.
The two situations pull against each other: minimizing the oscillation requires spikes with a small effect, but keeping the spike frequency low far from the target requires spikes with a large effect.
High spike frequencies are not a problem in theory, but depending on the actuator they may be undesirable in practice.
In power electronics, for example, the total switching loss grows with the switching frequency \cite[Ch.~4]{erickson_powerelectronics_2020}. 
This prompts an apparently obvious fix: if spike strength were proportional to the state, as in continuous control, would this not give asymptotic stability while softly bounding the spiking rate?
This kind of proportional, state-dependent transition is not unheard of in impulsive control \cite{briat_convexdwelltime_2013}, but, to the best of the authors' knowledge, the tools developed there do not fit the neuromorphic setting: they either ask the jump matrix to be strictly contracting \cite{hespanha_lyapunov_2008}, or they require satisfying discrete-time Lyapunov linear matrix inequalities on the flow--jump cycle for all possible dwell times \cite{briat_convexdwelltime_2013}.
Sections \ref{sec:initial_analysis} and \ref{sec:proof} detail why neither applies, in general, in the neuromorphic control setting.

The controller we study here keeps the integrate-and-fire behavior driven by the plant output, but makes its spike amplitude proportional to that output at firing time.
Because the spike now shrinks with the output, the limitation to practical stability disappears.
This only removes the first of the two problems: the spike frequency still grows far from the target, but each spike delivers a larger correction, so fewer are needed.
Although a graded-spike actuator might not be readily available, this work is a first step toward a design using an array of fixed-amplitude neurons working at different output scales.

Our first contribution is the formulation of what we call a \emph{single-input single-output (SISO) linear time-invariant (LTI) neuro-impulsive control system (NICS)}, and the proof that any such system built for a static output feedback (SOF) stabilizable plant is semi-globally asymptotically stable (sGAS), using the hybrid system formalism \cite{goebel_hybridsystems_2012} and an emulation approach \cite{nesic_emulation_2009, petri_emulationneuromorphic_2025}, in which the spiking controller is designed to emulate the closed-loop of a stabilizing continuous-time SOF.
The second contribution is an algorithm returning control parameters that make the NICS sGAS from a given set of initial conditions.
It is tested by stabilizing a linearized inverted pendulum at its upright equilibrium.

The paper is structured as follows. Section~\ref{sec:preliminaries} fixes notation, Section~\ref{sec:system_def} states the control problem, Section~\ref{sec:initial_analysis} develops the intermediate results, Section~\ref{sec:proof} presents the main result and its proof and derives an algorithm from it, Section~\ref{sec:example} illustrates the approach on a numerical example, and Section~\ref{sec:conclusion} concludes the paper.\ifwithappendix\ Appendix~\ref{appendix:proofs} contains the proofs of the intermediate results.\fi

All code and data are available on GitHub: \url{https://github.com/ChristianFernandezLorden/ECC27-FernandezLorden-SISONeuroImpulsive}.

\section{Preliminaries}\label{sec:preliminaries}

We denote by $\R$, $\mathbb{C}$, and $\N$ the sets of real, complex, and natural numbers, the latter including 0. 
Given a positive-definite matrix $P$, the energy norm of $\mathbf{x}$ is $\vnorm{\mathbf{x}}_P=\sqrt{\mathbf{x}^\top P\mathbf{x}}$, and $\mnorm{M}_P$ denote the matrix norm induced by it: $\mnorm{P^{1/2}MP^{-1/2}}_2$ for a square matrix $M$ and $\sqrt{CP^{-1}C^\top}$ for a row matrix $C$.
$\mu_P(M)$ denotes the logarithmic norm \cite{strom_logarithmicnorm_1975} of a square matrix $M$ induced by the $P$ energy vector norm. 
Lemmas holding for any vector norm are written without a subscript. 
The logarithmic norm property we use is $\mnorm{e^{At}}\le e^{\mu(A)t}\le e^{\mnorm{A}t}$.
For a symmetric $M$, $\lambda_\text{max}(M)$ and $\lambda_\text{min}(M)$ are its largest and smallest eigenvalues, and a square matrix is Hurwitz if all its eigenvalues have negative real part. 
By $W_0(x)$ we denote the principal branch of the Lambert $W$ function \cite{corless_lambertw_1996}, defined for $x\ge-1/e$.

We consider hybrid systems in the formalism of \cite{goebel_hybridsystems_2012}, meaning a system $\mathcal{H}$ is defined as
\begin{equation}
    \mathcal{H} : \begin{cases}
        \dot{\mathbf{z}} = f(\mathbf{z}), & \mathbf{z}\in\mathcal{C},\\
        \mathbf{z}^+ = g(\mathbf{z}), & \mathbf{z}\in\mathcal{D},
    \end{cases}\label{eq:hybrid_system}
\end{equation}
with $\mathbf{z}\in\R^n$ the state, $f$ and $\mathcal{C}$ the flow map and the flow set, and $g$ and $\mathcal{D}$ the jump map and the jump set. 
Hybrid time domains and solutions of \eqref{eq:hybrid_system} are defined as in \cite[Definitions 2.3 and 2.6]{goebel_hybridsystems_2012}, so each solution $\mathbf{z}$ has a domain $\dom \mathbf{z} = \bigcup_{i}[t_i, t_{i+1}]\times\{i\}$ with $0=t_0 \le t_1\le\cdots$, and we write $\mathbf{z}_i \eqdefsymbol \mathbf{z}(t_i,i)$ for the state after the $i$th jump ($\mathbf{z}_0$ being the initial condition).
We only consider maximal solutions in the sense of \cite[Definition~2.7]{goebel_hybridsystems_2012}.
A \emph{controlled hybrid system} is a system \eqref{eq:hybrid_system} whose data \cite[Definition~2.2]{goebel_hybridsystems_2012} are parametrized by $\theta\in\Theta$, the set of admissible control parameters.

\begin{definition}[Semi-global asymptotic stability]\label{def:semiglobal}
     A set $\mathcal{A}\subset\mathcal{U}$ of a controlled hybrid system with state space $\mathcal{U}$ and control parameters $\theta\in\Theta$ is said to be \emph{semi-globally asymptotically stable (sGAS)} if, for every compact set $\mathcal{Z}\subset\mathcal{U}$ of initial conditions, there exists $\theta\in\Theta$ such that $\mathcal{A}$ is pre-asymptotically stable from every $\mathbf{z}_0\in\mathcal{Z}$ in the sense of \cite[Definition~3.6]{goebel_hybridsystems_2012}.
\end{definition}

\section{Control system setup and problem statement}\label{sec:system_def}

We consider SISO LTI systems 
\begin{equation}
    \dot{\mathbf{x}} = A\mathbf{x} + Bu,\qquad y = C\mathbf{x},\label{eq:LTI}
\end{equation}
with scalar input $u(t)\in\R$, state $\mathbf{x}(t)\in\R^n$, and scalar output $y(t)\in\R$ at time $t\in\R_{\ge0}$.
We only consider systems \eqref{eq:LTI} that are unstable, i.e., there is one eigenvalue $\lambda_i$ of $A$ such that $\mathop{\mathrm{Re}}(\lambda_i)>0$, and satisfy the following assumption.

\begin{assumption}[Static output-feedback stabilizability]\label{assumption:hurwitz}
     The set $\mathcal{K}=\left\{K\in\mathbb{R}_{>0}:\text{$A-BKC$ is Hurwitz}\right\}$ is non-empty.\footnote{If the system is only stabilizable by a negative $K$, the sign of the feedback in \eqref{subeq:jump} can be swapped to regain a similar analysis.}
\end{assumption}
A necessary and sufficient condition for Assumption~\ref{assumption:hurwitz} is given in \cite[Theorem~5]{astolfi_sofproof_2001}.


This paper considers the stabilization of the origin of an LTI system \eqref{eq:LTI} using a feedback hybrid controller inspired by neural dynamics, and aims to show that any SISO LTI system stabilizable by SOF can also be stabilized by it.
Consider the hybrid control system \eqref{eq:hybrid_system} with data
\begin{subequations}\label{eq:lin_sys_prop_spike}
    \begin{align} \!\!\!\!f\left(\begin{pmatrix}\mathbf{x}\\\xi\end{pmatrix}\right)&\eqdefsymbol \begin{pmatrix}A\mathbf{x}\\|C\mathbf{x}|+I_b\end{pmatrix},&&\mathcal{C}\eqdefsymbol \R^n\times[0,\Delta],\label{subeq:lin_sys_prop_spike_flow}\\
    \!\!\!\!g\left(\begin{pmatrix}\mathbf{x}\\\xi\end{pmatrix}\right)&\eqdefsymbol \begin{pmatrix}(I-\alpha BC)\mathbf{x}\\0\end{pmatrix},&&\mathcal{D}\eqdefsymbol \R^n\times[\Delta,\infty),\label{subeq:jump}
    \end{align}
\end{subequations}
where $\xi\in\R_{\ge0}$ is the neuron state, $\Delta\in\R_{>0}$ the firing threshold, $I_b\in\R_{>0}$ a constant bias current, and $\alpha\in\R_{>0}$ the spike gain.
$\Delta$, $I_b$, and $\alpha$ are the design parameters of \eqref{eq:lin_sys_prop_spike}.
Observe that \eqref{eq:lin_sys_prop_spike} is the feedback interconnection of \eqref{eq:LTI} and a SISO neuro-impulsive controller with state $\xi$ that generates a control impulse proportional to $y=C\mathbf{x}$ each time the neural threshold is reached.
We call this closed loop a \emph{SISO LTI neuro-impulsive control system (NICS)}.

The flow of $\xi$ in \eqref{subeq:lin_sys_prop_spike_flow} is inspired by the well-known integrate-and-fire dynamics \cite{burkitt_ifreview_2006}, already used in a hybrid closed-loop model in \cite{petri_lifhybridanalysis_2024, petri_emulationneuromorphic_2025}, but the jump \eqref{subeq:jump} replaces the usual fixed-amplitude jump of the plant state with a jump proportional to the plant output at firing time, like a Dirac impulse sent as input to \eqref{eq:LTI}.
With fixed-amplitude spikes \cite[Section~III]{petri_emulationneuromorphic_2025}, a control input taking both signs requires at least two neurons, one per sign, whereas \eqref{subeq:jump} already carries the sign of $y$, so one neuron driven by the nonnegative input $|C\mathbf{x}|+I_b$ suffices.

Such a system has a unique maximal solution from every initial condition $(\mathbf{x}_0,\xi_0)\in\mathcal{C}\cup\mathcal{D}=\R^n\times\R_{\ge0}$\ifwithappendix\ (see Appendix~\ref{appendix:proofs})\fi.
Since $\dot{\xi}\ge I_b$ during flows and $\xi$ is reset to $0$ at each jump, this solution jumps infinitely many times. 
The problem we address is to design, for a system \eqref{eq:LTI} satisfying Assumption~\ref{assumption:hurwitz} and a compact set $\mathcal{Z}\subset\R^n\times\R_{\ge0}$ of initial conditions, the parameters $\Delta$, $I_b$, and $\alpha$ such that the set $\{\mathbf{0}\}\times[0,\Delta]$ is asymptotically stable for \eqref{eq:lin_sys_prop_spike} with a basin of attraction containing $\mathcal{Z}$.
In the rest of the paper, NICS is a shorthand for SISO LTI NICS and $\mathbf{z}\eqdefsymbol (\mathbf{x}, \xi)$ denotes the state of \eqref{eq:lin_sys_prop_spike}. 

\begin{remark}
    The bias current $I_b$ ensures an upper bound $\Delta/I_b$ on the time between two consecutive spikes, also called maximal dwell time, as shown in Lemma~\ref{lemma:dwell-time}.
    Asymptotic stability comes from the spike being proportional to the plant output, but our analysis also requires $I_b>0$.
    Since $|C\mathbf{x}|$ disappears from the trigger near the origin, without $I_b$, the time between spikes diverges to infinity as $\mathbf{x}_i\to\mathbf{0}$ and the effective gain $\alpha/\dwellt(\mathbf{x}_i)$ of Section~\ref{sec:initial_analysis} goes to zero.
\end{remark}

\section{Impact map analysis}\label{sec:initial_analysis}

Note that $BC$ is a rank-one matrix, so $I-\alpha BC$ must have 1 as an eigenvalue as soon as $n>1$, which means the jump map can never be a contraction.
Between jumps, $u=0$ and the plant follows its open-loop flow, which is not a contraction either since $A$ is not Hurwitz.
Stability can therefore only be proven from the combined effect of the flow and the jump, which rules out methods relying on contracting jump matrices, e.g., \cite{hespanha_lyapunov_2008}, and leads us to define the so-called impact map.

\begin{definition}[Impact map]\label{def:impact}
    The \emph{impact map} of a solution of a hybrid system $\mathcal{H}$ that satisfies the uniqueness conditions of \cite[Proposition 2.11]{goebel_hybridsystems_2012} is the function $\mathcal{F}:\mathbb{R}^n\rightarrow\mathbb{R}^n$ that maps the hybrid system state immediately after a jump to its state immediately after the next jump:
    \begin{equation}
        \mathbf{z}_{i+1} = \mathcal{F}(\mathbf{z}_{i})\quad\text{whenever }(t_{i+1},i+1)\in\dom \mathbf{z}.
    \end{equation}
\end{definition}

In NICS~\eqref{eq:lin_sys_prop_spike}, the impact map (Definition~\ref{def:impact}) of a solution is a linear flow followed by a linear jump, hence linear and, for any jump after the first, it is determined by a transition matrix  
\begin{equation}
    \impactmap(\dwellt_{\Delta,I_b}(\mathbf{x}),\alpha)\eqdefsymbol \begin{bmatrix}(I-\alpha BC)\,e^{A\dwellt_{\Delta,I_b}(\mathbf{x})} & \mathbf{0}_{n\times 1} \\ \mathbf{0}_{1\times n} & 0\end{bmatrix},\label{eq:transition_matrix}
\end{equation} 
with $\dwellt_{\Delta,I_b}(\mathbf{x}):\R^n\rightarrow\R_{>0}$ the time it takes $\xi$ to reach $\Delta$ from a post-jump state $\mathbf{z}=(\mathbf{x},0)$, i.e., the dwell time \cite[Section~2.4]{goebel_hybridsystems_2012} from $\mathbf{z}$.
The zeros on the bottom row represent the reset of $\xi$, and for brevity we denote the non-zero block by $\impactmap(\mathbf{x})$ and the dwell time $\dwellt_{\Delta,I_b}(\mathbf{x})$ by $\dwellt(\mathbf{x})$.

To first order in $\dwellt(\mathbf{x})$, the transition matrix $(I-\alpha BC)\,e^{A\dwellt(\mathbf{x})}$ agrees with the flow matrix $e^{(A-(\alpha/\dwellt(\mathbf{x})) BC)\dwellt(\mathbf{x})}$ of a continuous SOF loop with gain ${\alpha}/{\dwellt(\mathbf{x})}$, which Lemma~\ref{lemma:jump_emulation} makes quantitative.


In Lemmas~\ref{lemma:jump_emulation}--\ref{lemma:x0_decrease_dwell_time}, $\mathbf{z}$ denotes any maximal solution of \eqref{eq:lin_sys_prop_spike} and $i\in\N_{\ge1}$, so that $\xi_i=0$.
The first flow, which starts from an arbitrary $\xi_0$, is treated separately in the proof of Theorem~\ref{thm:main}.

\begin{lemma}[Continuous-feedback emulation]\label{lemma:jump_emulation}
    Let $i\in\N_{\ge1}$ and $K\in\R$, and define $\kappa(\mathbf{x}_i)\eqdefsymbol K- {\alpha}/{\dwellt(\mathbf{x}_i)}$.
    If $\dwellt(\mathbf{x}_i)<\min\bigl(\mnorm{A}^{-1},\mnorm{A-BKC}^{-1}\bigr)$, then
    \begin{multline}
        \mnorm{\impactmap(\mathbf{x}_i)}\le\mnorm{e^{(A-BKC)\dwellt(\mathbf{x}_i)}} +C_{1}(\kappa(\mathbf{x}_i))\dwellt(\mathbf{x}_i)\\+C_{2}(\kappa(\mathbf{x}_i)){\dwellt(\mathbf{x}_i)}^2+C_{3}(\kappa(\mathbf{x}_i)){\dwellt(\mathbf{x}_i)}^3,
    \end{multline}
    with 
    \begin{subequations}\label{eq:constants}
    \begin{gather}
        C_{1}(\kappa)\eqdefsymbol |\kappa|\mnorm{BC},\\
        C_{2}(\kappa)\eqdefsymbol \mnorm{BC}\!\left(\!\left(|K|+|\kappa|\right)\mnorm{A}+\frac{K^2}{2}|CB|\right),\\ 
        \begin{aligned}C_{3}(\kappa)\eqdefsymbol &\bigl(\mnorm{A}^3+\mnorm{A-BKC}^3\bigr)\frac{2e-5}{2}\\&+\bigl(|K|+|\kappa|\bigr)\mnorm{BC}\mnorm{A}^2\,(e-2).
        \end{aligned}
    \end{gather}
    \end{subequations}
\end{lemma}

Therefore, for any solution of \eqref{eq:lin_sys_prop_spike}, the ratio ${\alpha}/{\dwellt(\mathbf{x})}$ is the effective static output-feedback gain that the spiking controller emulates when flowing from a state $\mathbf{z}=(\mathbf{x},0)$ and jumping, and $\kappa(\mathbf{x})$ is the deviation of that effective gain from the reference gain $K$.
Note that this ratio is not a design parameter: $\alpha$ is, but $\dwellt(\mathbf{x})$ is determined by the plant dynamics through the firing condition.
Since $\kappa(\mathbf{x})$ is the coefficient of the first-order term of the difference between the transition matrix and the continuous flow, the two are close for small $\dwellt(\mathbf{x})$ when $|\kappa(\mathbf{x})|$ is small.

For Hurwitz $A-BKC$ and a positive-definite $Q^*\in\R^{n\times n}$, the Lyapunov equation 
\begin{equation}
    (A-BKC)^\top P^*+P^*(A-BKC)=-Q^* \label{eq:lyapunov}
\end{equation}
has a unique positive-definite solution $P^*$, which defines the Lyapunov function $V(\mathbf{x})=\mathbf{x}^\top P^*\mathbf{x}$ of the continuous loop $\dot{\mathbf{x}}=(A-BKC)\mathbf{x}$.
The standard Lyapunov decay estimate then guarantees that $\mnorm{e^{(A-BKC)\dwellt}}_{P^*}\le e^{-\lambda_0\dwellt/2}$, with $\lambda_0\eqdefsymbol\lambda_\text{min}(Q^*)/\lambda_\text{max}(P^*)>0$ \cite[Section~8.5]{hespanha_linear_2018}.
Lemma~\ref{lemma:jump_emulation} expressed in the $P^*$-norm then bounds $\mnorm{\impactmap(\mathbf{x})}_{P^*}$ by $e^{-\lambda_0\dwellt(\mathbf{x})/2}+\sum_{k=1}^{3}C_k(\kappa(\mathbf{x}))\,\dwellt(\mathbf{x})^k$.
As Lemma~\ref{lemma:contractive_range} shows, under the right conditions this bound is smaller than 1, which means that the transition matrix \eqref{eq:transition_matrix} contracts $\mathbf{x}$.

\begin{lemma}[Existence of a contracting $(\alpha,\dwellt(\mathbf{x}))$ range]\label{lemma:contractive_range}
    Suppose Assumption~\ref{assumption:hurwitz} is satisfied, let $K\in\mathcal{K}$, let $Q^*\in\mathbb{R}^{n\times n}$ be positive definite, and let $P^*$ be the solution of the Lyapunov equation \eqref{eq:lyapunov} and $\lambda_0\eqdefsymbol\lambda_\text{min}(Q^*)/\lambda_\text{max}(P^*)>0$.
    Let $i\in\N_{\ge1}$ and assume $\dwellt(\mathbf{x}_i)<\min\bigl(\mnorm{A}_{P^*}^{-1},\mnorm{A-BKC}_{P^*}^{-1}\bigr)$.
    Suppose the following holds for at least one $\overline\kappa\in\bigl(0,\lambda_0/(2\mnorm{BC}_{P^*})\bigr)$:
    \begin{gather}
        \!\!\!\!\dwellt(\mathbf{x}_i)<\frac{-(\overline{C_2}+\frac{\lambda_0^2}{8})+\sqrt{(\overline{C_2}+\frac{\lambda_0^2}{8})^2+4\overline{C_3}\bigl(\frac{\lambda_0}{2}-\overline{C_1}\bigr)}}{2\overline{C_3}},\label{eq:contractive_dwell_time_cond}\\
        |\kappa(\mathbf{x}_i)|<\overline\kappa.\label{eq:kappa_bound}
    \end{gather}
    Then $\mnorm{\impactmap(\mathbf{x}_i)}_{P^*}\le g(\dwellt(\mathbf{x}_i))<1$, where $g(\dwellt)\eqdefsymbol e^{-\lambda_0 \dwellt/2}+\overline{C_1}\dwellt+\overline{C_2}\dwellt^2+\overline{C_3}\dwellt^3<1$ for every $\dwellt>0$ satisfying \eqref{eq:contractive_dwell_time_cond}. 
    Here $\overline{C_1}\eqdefsymbol C_1(\overline\kappa)$, $\overline{C_2}\eqdefsymbol C_2(\overline\kappa)$, and $\overline{C_3}\eqdefsymbol C_3(\overline\kappa)$ are the constants \eqref{eq:constants} of Lemma~\ref{lemma:jump_emulation} expressed in the $P^*$-norm.
\end{lemma}


$\kappa(\mathbf{x})$ and $\dwellt(\mathbf{x})$ are not controller parameters.
But $\kappa(\mathbf{x})$ depends on the dynamics only through $\dwellt(\mathbf{x})$, and since Lemma~\ref{lemma:jump_emulation} and \eqref{eq:contractive_dwell_time_cond} impose upper bounds on $\dwellt(\mathbf{x})$ while \eqref{eq:kappa_bound} bounds $\alpha/\dwellt(\mathbf{x})$ to a range, it suffices to bound $\dwellt(\mathbf{x})$ in terms of the controller parameters, which the next lemma does.


\begin{lemma}[Per-jump dwell time bounds]\label{lemma:dwell-time}
    For any $i\in\N_{\ge1}$,
    \begin{equation}
        0<\underline{\dwellt}(\mathbf{x}_i)\le \dwellt(\mathbf{x}_i) \le\overline{\dwellt},
    \end{equation}
    with 
    \begin{equation}
        \overline{\dwellt}\eqdefsymbol\frac{\Delta}{I_b}\label{eq:dwell_time_upper_bound}
    \end{equation}
    and 
    \begin{equation}
       \!\!\!\underline{\dwellt}(\mathbf{x}_i)\eqdefsymbol\frac{1}{\mu(A)}\ln\left(\frac{ W_0\!\left(\frac{\mnorm{C}\,\vnorm{\mathbf{x}_i}}{I_b}\,e^{\frac{\mnorm{C}\,\vnorm{\mathbf{x}_i}}{I_b}+\frac{\Delta\mu(A)}{I_b}}\right)}{\frac{\mnorm{C}\,\vnorm{\mathbf{x}_i}}{I_b}}\right).\label{eq:dwell_time_lower_bound}
    \end{equation}
\end{lemma}

The upper bound \eqref{eq:dwell_time_upper_bound} alone is enough for Lemma~\ref{lemma:jump_emulation} and condition \eqref{eq:contractive_dwell_time_cond}, as both only require $\dwellt(\mathbf{x})$ to be small enough.
Condition \eqref{eq:kappa_bound}, however, requires $\alpha/\dwellt(\mathbf{x})$ to lie in $(K-\overline\kappa, K+\overline\kappa)$ for every possible $\dwellt(\mathbf{x})$, which an upper bound cannot give.
Multiplying the range by $\dwellt(\mathbf{x})$, the condition becomes finding a single $\alpha$ between two linear functions of $\dwellt(\mathbf{x})$ over the interval of Lemma~\ref{lemma:dwell-time}.
Its existence is not trivial. 
The next lemma gives a sufficient condition for existence using both bounds of Lemma~\ref{lemma:dwell-time}.

\begin{lemma}[Sufficient condition on $I_b$ and $\Delta$ for an admissible ${\alpha}/{\dwellt(\mathbf{x})}$ range]\label{lemma:alpha_range_existence}
    Let $0<a<b$ and $i\in\N_{\ge1}$, and suppose the following holds
    \begin{equation}
        I_b>\Delta\,\frac{(a/b)\,\mu(A)}{\ln\!\left(\dfrac{\dfrac{(a/b)\Delta\mu(A)}{\mnorm{C}\vnorm{\mathbf{x}_i}}}{W_0\!\left(\dfrac{(a/b)\Delta\mu(A)}{\mnorm{C}\vnorm{\mathbf{x}_i}}\,e^{\frac{(a/b)\Delta\mu(A)}{\mnorm{C}\vnorm{\mathbf{x}_i}+(1-a/b)\Delta\mu(A)}}\right)}\right)}.\label{eq:range_I0_cond}
    \end{equation}
    Then there exists $\alpha\in\R_{>0}$ such that $a<\alpha/\dwellt<b$ for all $\dwellt\in[\underline{\dwellt}(\mathbf{x}_i),\overline{\dwellt}]$.
\end{lemma}

Suppose the parameters of \eqref{eq:lin_sys_prop_spike} are chosen so that $\impactmap(\mathbf{x}_i)$ is contractive, i.e., $\vnorm{\mathbf{x}_i}_{P^*}>\vnorm{\mathbf{x}_{i+1}}_{P^*}$.
The lower bound \eqref{eq:dwell_time_lower_bound} depends on $\vnorm{\mathbf{x}}$, and this dependency propagates to Lemma~\ref{lemma:alpha_range_existence}, which becomes harder to satisfy as that bound shrinks.
So parameters making $\impactmap(\mathbf{x}_i)$ contractive also make the next transition contractive if the bound is monotonically decreasing in $\vnorm{\mathbf{x}}$.


\begin{lemma}[Dwell time lower bound monotonicity]\label{lemma:x0_decrease_dwell_time}
    Let $\Delta,I_b\in\R_{>0}$. The lower bound $\underline{\dwellt}(\mathbf{x})$ defined by \eqref{eq:dwell_time_lower_bound} depends on $\mathbf{x}\in\R^n\setminus\{\mathbf{0}\}$ only through $\vnorm{\mathbf{x}}$, and is a continuous, strictly decreasing function of $\vnorm{\mathbf{x}}\in\R_{>0}$.
\end{lemma}

In Lemma~\ref{lemma:alpha_range_existence}, $I_b$ is easy to isolate but $\Delta$ is not, and the dwell time conditions of Lemmas~\ref{lemma:jump_emulation} and \ref{lemma:contractive_range} are met once they hold at $\overline{\dwellt}=\Delta/I_b$, which motivates a common $I_b$ lower bound.

\begin{definition}[$I_b$ total lower bound]\label{def:I0_total_bound}
    Suppose Assumption~\ref{assumption:hurwitz} is satisfied.
    Given $\mathbf{x}$, a gain $K\in\mathcal{K}$, a positive-definite $Q^*\in \mathbb{R}^{n\times n}$, a threshold $\Delta\in\R_{>0}$, and $\overline{\kappa}\in\R_{>0}$, the value $\underline{I_{b}}(\mathbf{x}, K, Q^*, \Delta, \overline{\kappa})$, denoted $\underline{I_{b}}(\mathbf{x})$ for brevity, is the infimum of the $I_b\in\R_{>0}$ simultaneously satisfying the condition of Lemma~\ref{lemma:alpha_range_existence} with $a=K-\overline{\kappa}$ and $b=K+\overline{\kappa}$ (which satisfy\footnote{$K-\overline{\kappa}$ is always greater than 0 for unstable systems.  
    Otherwise, one could choose $\kappa(\mathbf{x}) = K$ in Lemma~\ref{lemma:contractive_range}, which would imply $\mnorm{e^{A\dwellt(\mathbf{x})}}_{P^*}<1$ for some $\dwellt(\mathbf{x})$, which is impossible if the system is unstable.} $0<a<b$) and the $\dwellt(\mathbf{x})$ conditions of Lemmas~\ref{lemma:jump_emulation} and \ref{lemma:contractive_range} at the worst case $\dwellt(\mathbf{x})=\overline{\dwellt}=\Delta/I_b$, all in the energy norm of Lemma~\ref{lemma:contractive_range}.
\end{definition}
Each of these conditions is a strict lower bound on $I_b$, so all of them hold if and only if $I_b>\underline{I_b}(\mathbf{x})$.

\section{Stability of NICSs}\label{sec:proof}

In the previous section we showed that the impact map \eqref{eq:transition_matrix} contracts the LTI state $\mathbf{x}$ when the dwell time and the effective gain satisfy Lemma~\ref{lemma:contractive_range}.
Definition~\ref{def:I0_total_bound}, built on Lemmas~\ref{lemma:dwell-time} and \ref{lemma:alpha_range_existence}, turns these into conditions on the controller parameters. 
This is enough to show that suitable $I_b$ and $\alpha$ exist for every $\Delta>0$ and every compact set of initial conditions such that the set $\{\mathbf{0}\}\times[0, \Delta]$ is asymptotically stable and attracts all solutions starting in it.

\begin{theorem}[Existence of a stabilizing spike controller for static-output-feedback-stabilizable SISO linear systems]\label{thm:main}
    Consider a NICS~\eqref{eq:lin_sys_prop_spike} for which Assumption~\ref{assumption:hurwitz} holds.
    Then, for every $\Delta>0$, the set $\{\mathbf{0}\}\times[0, \Delta]$ is sGAS in the sense of Definition~\ref{def:semiglobal} with $\theta=(I_b,\alpha)$ and $\Theta=\R_{>0}^2$.
\end{theorem}

\begin{proof}
    Let $K\in\mathcal{K}$ and let $Q^*\in\mathbb{R}^{n\times n}$ be positive definite.
    Then the Lyapunov equation \eqref{eq:lyapunov} has a unique positive-definite solution $P^*\in\mathbb{R}^{n\times n}$, and $Q^*$ and $P^*$ define $\lambda_0\eqdefsymbol\lambda_\text{min}(Q^*)/\lambda_\text{max}(P^*)>0$ (see Section~\ref{sec:initial_analysis}).
    Let $\Delta>0$ and $\overline\kappa\in\bigl(0,\lambda_0/(2\mnorm{BC}_{P^*})\bigr)$.
    For any compact set $\mathcal{Z}\subset\R^n\times\R_{\ge0}$ of initial conditions, let $\hat{\mathbf{x}}\in\operatorname*{arg\,max}_{\mathbf{x}\,:\,(\mathbf{x},\xi)\in\mathcal{Z}}\vnorm{\mathbf{x}}_{P^*}$ be the worst-case initial condition, where we assume without loss of generality that $\hat{\mathbf{x}}\neq\mathbf{0}$.
    Every norm below is in the $P^*$-norm, and $\mu\eqdefsymbol\mu_{P^*}(A)>0$ since \eqref{eq:LTI} is unstable.

    Since the system may start with $\xi_0\neq0$, the first flow can be arbitrarily short and the first jump must be treated separately.
    Let $I_b>0$ and $\alpha\in\bigl(0,(K+\overline{\kappa})\overline{\dwellt}\bigr)$, with $\overline{\dwellt}=\Delta/I_b$ as in \eqref{eq:dwell_time_upper_bound}, and let $\mathbf{z}=(\mathbf{x},\xi)$ be a maximal solution of the NICS~\eqref{eq:lin_sys_prop_spike} from $\mathbf{z}_0=(\mathbf{x}_0,\xi_0)$.
    Its first jump occurs at a time $t_1\in[0,\overline{\dwellt}]$, since $t_1=0$ if $\xi_0\ge\Delta$, as the system cannot flow from such a state, and $t_1\le(\Delta-\xi_0)/I_b\le\overline{\dwellt}$ if $\xi_0<\Delta$.
    After this jump, the neuron state is reset to $0$ and the LTI state is $\mathbf{x}_1=(I-\alpha BC)e^{At_1}\mathbf{x}_0$.
    Using $\mnorm{I-\alpha BC}\le1+\alpha\mnorm{BC}$ and the logarithmic norm property $\mnorm{e^{At_1}}\le e^{\mu t_1}\le e^{\mu\overline{\dwellt}}$, we have $\vnorm{\mathbf{x}_1}\le c(I_b)\vnorm{\mathbf{x}_0}$ with
    \begin{equation}
        c(I_b)\eqdefsymbol\Bigl(1+(K+\overline{\kappa})\mnorm{BC}\tfrac{\Delta}{I_b}\Bigr)e^{\mu\frac{\Delta}{I_b}}.\label{eq:first_jump_factor}
    \end{equation}
    Both factors are at least 1 and non-increasing in $I_b$, so $c\ge1$ and $c$ is non-increasing.
    
    The first jump can leave the set of initial conditions, so the parameters must be chosen for an enlarged set.
    Let $I_{b,0}\eqdefsymbol\underline{I_b}(\hat{\mathbf{x}})$ be the bound of Definition~\ref{def:I0_total_bound} at the worst-case initial condition, let $\overline{c}\eqdefsymbol c(I_{b,0})$, and let $\bar{\mathbf{x}}\in\R^n$ be any state with $\vnorm{\bar{\mathbf{x}}}=\overline{c}\vnorm{\hat{\mathbf{x}}}$.
    The parameters are then
    \begin{equation}
        I_b>\max\bigl(I_{b,0},\underline{I_b}(\bar{\mathbf{x}})\bigr),\quad \alpha\in\bigl((K-\overline{\kappa})\overline{\dwellt},(K+\overline{\kappa})\underline{\dwellt}(\bar{\mathbf{x}})\bigr).\label{eq:parameter_choice}
    \end{equation}
    The range of $\alpha$ is non-empty since the condition of Lemma~\ref{lemma:alpha_range_existence} holds at $\bar{\mathbf{x}}$, and every $\alpha$ in it satisfies $0<\alpha<(K+\overline{\kappa})\overline{\dwellt}$ since $\underline{\dwellt}(\bar{\mathbf{x}})\le\overline{\dwellt}$, so the bound on the first jump applies.
    Since $I_b>I_{b,0}$ and $c$ is non-increasing, $c(I_b)\le\overline{c}$, so $\vnorm{\mathbf{x}_1}\le\overline{c}\vnorm{\mathbf{x}_0}\le\vnorm{\bar{\mathbf{x}}}$ for every initial condition with $\vnorm{\mathbf{x}_0}\le\vnorm{\hat{\mathbf{x}}}$ and $\xi_0\ge0$, which includes every $\mathbf{z}_0\in\mathcal{Z}$.

    Now consider the solutions $\mathbf{z}$ of the NICS~\eqref{eq:lin_sys_prop_spike} with the parameters \eqref{eq:parameter_choice}.
    After the first jump, every post-jump state is of the form $(\mathbf{x}_i,0)$ with $i\in\N_{\ge1}$, so the transition matrix \eqref{eq:transition_matrix} describes the evolution of $\mathbf{x}_i$.
    Let $g$ be the function of Lemma~\ref{lemma:contractive_range}.
    Since $g$ is convex on $\R_{\ge0}$, its maximum over $[\underline{\dwellt}(\bar{\mathbf{x}}),\overline{\dwellt}]$ is reached at an endpoint, and both endpoints satisfy \eqref{eq:contractive_dwell_time_cond} since $\overline{\dwellt}$ does, so
    \begin{equation}
        \overline{\gamma}\eqdefsymbol\max\bigl(g(\underline{\dwellt}(\bar{\mathbf{x}})),\,g(\overline{\dwellt})\bigr)<1 .\label{eq:contraction_rate}
    \end{equation}
    Suppose $\vnorm{\mathbf{x}_i}\le\vnorm{\bar{\mathbf{x}}}$ for some $i\in\N_{\ge1}$.
    If $\mathbf{x}_i=\mathbf{0}$, the LTI state stays at $\mathbf{0}$.
    Otherwise, Lemmas~\ref{lemma:dwell-time} and~\ref{lemma:x0_decrease_dwell_time} give $\dwellt(\mathbf{x}_i)\in[\underline{\dwellt}(\bar{\mathbf{x}}),\overline{\dwellt}]$, so $|\kappa(\mathbf{x}_i)|<\overline{\kappa}$ by the choice of $\alpha$, and the dwell time conditions of Lemmas~\ref{lemma:jump_emulation} and~\ref{lemma:contractive_range} hold since they hold at $\overline{\dwellt}$ by Definition~\ref{def:I0_total_bound}.
    Lemma~\ref{lemma:contractive_range} then bounds the norm of the transition matrix by $\mnorm{\impactmap(\mathbf{x}_i)}\le g(\dwellt(\mathbf{x}_i))\le\overline{\gamma}$, so $\vnorm{\mathbf{x}_{i+1}}\le\overline{\gamma}\vnorm{\mathbf{x}_i}<\vnorm{\bar{\mathbf{x}}}$.
    Since $\vnorm{\mathbf{x}_1}\le\vnorm{\bar{\mathbf{x}}}$, repeating this argument gives, for every $i\in\N_{\ge1}$,
    \begin{equation}
        \vnorm{\mathbf{x}_i}\le\overline{\gamma}^{\,i-1}\vnorm{\mathbf{x}_1}\le\overline{\gamma}^{\,i-1}\,\overline{c}\vnorm{\mathbf{x}_0}.
    \end{equation}
    Since the system can only flow for a maximum of $\overline{\dwellt}$, the maximum norm it can attain from a state $\mathbf{x}_i$ is $e^{\mu\overline{\dwellt}}\vnorm{\mathbf{x}_i}$.
    Since $\overline{c}\ge1$, it follows that, for every $(t,i)\in\dom\mathbf{z}$,
    \begin{equation}
        \vnorm{\mathbf{x}(t,i)}\le\overline{c}\,e^{\mu\overline{\dwellt}}\,\overline{\gamma}^{\,\max(i-1,0)}\vnorm{\mathbf{x}_0}.\label{eq:state_bound}
    \end{equation}
    Note also that every maximal solution is complete and non-Zeno, since $0<\underline{\dwellt}(\bar{\mathbf{x}})\le\dwellt(\mathbf{x}_i)\le\overline{\dwellt}$ for every $i\in\N_{\ge1}$, and that $t\le(i+1)\overline{\dwellt}$ for every $(t,i)\in\dom\mathbf{z}$.

    Attractivity follows since $t+i\rightarrow\infty$ implies $i\rightarrow\infty$, so \eqref{eq:state_bound} with $\overline{\gamma}<1$ yields $\lim_{t+i\rightarrow\infty}\vnorm{\mathbf{x}(t,i)}=0$ and thus $\lim_{t+i\rightarrow\infty}\mathbf{x}(t,i)=\mathbf{0}$.
    Stability follows since, for every $\varepsilon>0$, $\delta\eqdefsymbol\min\bigl(\vnorm{\hat{\mathbf{x}}},\varepsilon\,e^{-\mu\overline{\dwellt}}/\overline{c}\bigr)$ is such that $\vnorm{\mathbf{x}_0}\le\delta$, with any $\xi_0\ge0$, implies $\vnorm{\mathbf{x}(t,i)}\le\varepsilon$ for every $(t,i)\in\dom\mathbf{z}$ by \eqref{eq:state_bound}.

    Finally, the neuron state $\xi$ is always contained in $[0,\Delta]$, except possibly at $t=0$, so asymptotic stability of $\mathbf{0}$ for $\mathbf{x}$ gives asymptotic stability of the set $\{\mathbf{0}\}\times[0,\Delta]$ for $\mathbf{z}$.
\end{proof}
The Theorem~shows that parameters always exist for which the LTI state contracts at each flow--jump cycle by emulating a SOF controller, and thus converges to $\mathbf{0}$.
Since $\xi$ stays in $[0, \Delta]$ and does not count for stability, the attractor is the set $\{\mathbf{0}\}\times[0, \Delta]$ rather than a point.




\begin{figure}
    \centering
    \noindent\hrule
    \begin{algorithmic}[1]
    \Require A system model as in \eqref{eq:LTI}, a compact set of initial conditions $\mathcal{Z}$, and a gain $K$ such that Assumption~\ref{assumption:hurwitz} holds. 
    \State Choose a positive-definite $Q^*$ and compute $P^*$ using the Lyapunov continuous-time equation. 
    \State Find $\hat{\mathbf{x}}\in\operatorname*{arg\,max}_{\mathbf{x}\,:\,(\mathbf{x},\xi)\in\mathcal{Z}}\vnorm{\mathbf{x}}_{P^*}$. 
    \State Choose $\Delta>0$. 
    \State Choose a $\overline{\kappa}\in\bigl(0,\lambda_0/(2\mnorm{BC}_{P^*})\bigr)$. 
    \State Choose $I_b > \underline{I_b}(\bar{\mathbf{x}})$, with $\bar{\mathbf{x}}$ being any state that satisfies $\vnorm{\bar{\mathbf{x}}}_{P^*}=c(\underline{I_b}(\hat{\mathbf{x}}))\vnorm{\hat{\mathbf{x}}}_{P^*}$, with $c$ as defined in \eqref{eq:first_jump_factor}.
    \State Compute the minimum and maximum dwell times $\underline{\dwellt}(\bar{\mathbf{x}})$ and $\overline{\dwellt}$ using Lemma~\ref{lemma:dwell-time}.
    \State Choose $\alpha \in\bigl((K-\overline{\kappa})\overline{\dwellt},(K+\overline{\kappa})\underline{\dwellt}(\bar{\mathbf{x}})\bigr)$.
    \end{algorithmic}
    \noindent\hrule
    \caption{Algorithm to design a stable NICS from a stabilizing gain and a set of initial conditions. In Section~\ref{sec:example} we take $Q^*,P^*$ maximizing $\lambda_0$ by convex optimization (Julia, \textsc{Convex} and \textsc{SCS}), $\Delta=\mnorm{C}_{P^*}\vnorm{\hat{\mathbf{x}}}_{P^*}e^{\mu_{P^*}(A)}$, $\overline{\kappa}$ minimizing $\underline{I_b}(\bar{\mathbf{x}})$, $I_b=1.5\underline{I_b}(\bar{\mathbf{x}})$, and $\alpha$ in the middle of its range.}
    \label{fig:algo}
\end{figure}

The proof is constructive: Fig.~\ref{fig:algo} presents an algorithm based on it, designing the controller from the plant parameters and a set of initial conditions. The algorithm chooses six parameters: the controller parameters ($\Delta$, $I_b$, $\alpha$) and three intermediate ones ($Q^*$, $P^*$, $\overline{\kappa}$) that follow from the emulation approach, $Q^*$ and $P^*$ coming from the Lyapunov equation of the emulated continuous system and $\overline{\kappa}$ being the maximum deviation of the effective gain.
Interestingly, the choice of $\Delta$ is completely free: it may be optimized against some objective, but a stabilizing controller exists for every $\Delta>0$. 


One may ask whether stability could be established with methods relying on parameter-dependent LMIs such as \cite{briat_convexdwelltime_2013}.
With the jump map constrained to the form $I-\alpha BC$, such results can certify a given set of parameters, but they cannot directly prove that stabilizing $(\alpha,\Delta,I_b)$ exist for the whole class of SOF stabilizable SISO systems.

One may also wonder why the term $|C\mathbf{x}|$ is included in \eqref{subeq:lin_sys_prop_spike_flow}.
If it is removed, firing becomes periodic, and global stability can be proven with Lemmas~\ref{lemma:jump_emulation} and~\ref{lemma:contractive_range} (periodic case already covered by \cite[Theorem~2.1]{briat_convexdwelltime_2013}).
The output-driven firing, i.e., the integrate-and-fire mechanism, may have two benefits: faster convergence than periodic firing and, when $\mathcal{K}=(K_{\min},\infty)$ and $|1-\alpha CB|\le1$, bounded trajectories for all $\Delta$ and $I_b$.

\section{Numerical example}\label{sec:example}

We consider an inverted pendulum linearized around its upright equilibrium, with a simplified motor model.
It is a SISO LTI system with state $\mathbf{x} = (\theta, \dot{\theta}, I) \in \R^3$, input $u\in\R$, and matrices
\begin{equation}
    A = \begin{bmatrix}0 & 1 & 0\\p_1 & -p_2 &p_3\\ 0  & -p_4 & -p_5\end{bmatrix},\quad B=\begin{bmatrix}0\\0\\p_6\end{bmatrix},\quad C=\begin{bmatrix}1&0 &0\end{bmatrix},
\end{equation}
with $p_1,p_2,p_3,p_4,p_5,p_6\in \R_{>0}$.


We consider $p_1=1$, $p_2=2$, $p_3=1$, $p_4=1$, $p_5=2$, and $p_6=1$.
Assumption~\ref{assumption:hurwitz} holds with $\mathcal{K}=(2,18)$.
We choose $K = 2.8$ for our design.
Designing for a compact set $\mathcal{Z}$ whose largest element in the $P^*$-norm is $(0.5,0.5,0)$ and applying the algorithm outlined in Fig.~\ref{fig:algo} with the choices in the caption yields $I_b = \num{1251.82}$, $\Delta = \num{3.55}$, and $\alpha =\num{0.00794}$.

\begin{figure}
    \centering
    \includegraphics{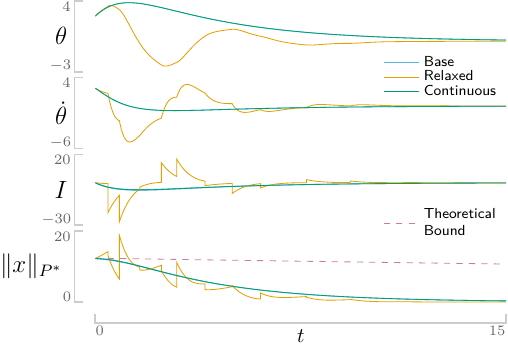}
    \caption{Trajectories of the controlled linearized inverted pendulum from $(\theta,\dot\theta, I) = (2.5,2.5,0.0)$, for the parameters returned by the algorithm (blue), the relaxed parameters of Section~\ref{sec:example} (red), and the emulated continuous loop (green). From top to bottom: $\theta$, $\dot\theta$, $I$, and the $P^*$-energy norm of the state, for which the theoretical minimum decay is also plotted.}
    \label{fig:example}

\end{figure}

Fig.~\ref{fig:example} compares the closed loop under the parameters returned by the algorithm with a ``relaxed'' choice that keeps ${\alpha I_b}/{\Delta}$ constant but divides $I_b$ by 1500 and $\Delta$ by 2, and with the continuous loop being emulated.
The relaxed controller emulates the same gain at the equilibrium but no longer fits the bounds established by \eqref{eq:contraction_rate}.
Since $I_b$ is reduced much more than $\Delta$, it achieves a far larger maximum dwell time, lowering the frequency of actuation at the origin by a factor of $\num{750}$.

Two observations stand out.
First, the algorithm's parameters yield trajectories closely following the continuous system, as expected from an emulation approach, whereas the relaxed controller diverges from it quickly even though it emulates the same gain at the origin.
Second, the controller can sit outside the bounds fixed by the proof and still stabilize the plant: the original one starts five times outside its design set, and the relaxed one also uses parameters beyond the bounds of Theorem~\ref{thm:main}.
The returned parameters are therefore only sufficient, and emulating a single gain $K$ ignores the whole range of stabilizing gains the plant admits.

\section{Conclusion}\label{sec:conclusion}

We have designed a controller at the crossroads of neuromorphic and traditional impulsive control, using a single integrate-and-fire-like neural variable with a spike amplitude proportional to the output, and shown that one such neuron semi-globally asymptotically stabilizes any SOF stabilizable SISO LTI system.
We also proposed an algorithm to design the controller from the plant parameters, and tested it in simulations on a linearized inverted pendulum.

At this stage the contribution is mostly conceptual: an actuator that can deliver pulses of any amplitude can usually deliver a continuous signal as well, and the proportional spike then gives up the main practical advantage of neuromorphic control.
The result is still useful, as it is the shrinking of the spikes with the state, and not only the firing time, that decides between asymptotic and practical stability.
This can still be used with fixed-amplitude spikes by building arrays of neurons working at different scales so that the total correction is quantized but keeps decreasing with the state, much like the motor units recruited in biological muscle depending on the required torque \cite{fuglevand_motorunits_1993}; such an array should converge with fewer spikes and a smaller ultimate bound than a single neuron pair.
The work could also be extended to the multiple-input multiple-output case, which would probably require average dwell times, and it would be interesting to see how much of the conservatism of the proof and of the algorithm can be pruned.

\ifwithappendix\appendix\section{Appendix} 

\subsection{Proofs}\label{appendix:proofs}

\begin{proof}[Proof of uniqueness of solutions of NICSs \eqref{eq:lin_sys_prop_spike}]
    Points (a) and (c) of \cite[Proposition 2.11]{goebel_hybridsystems_2012} hold for NICSs since the flow map \eqref{subeq:lin_sys_prop_spike_flow} is globally Lipschitz, so solutions of the flow are unique \cite[Theorem~3.2]{sideris_ode_2013}, and the jump map \eqref{subeq:jump} is single-valued.

    Point (b) of \cite[Proposition 2.11]{goebel_hybridsystems_2012} only concerns the states in $\mathcal{C}\cap\mathcal{D}$, i.e., $\mathbf{z}=(\mathbf{x},\Delta)$ with $\mathbf{x}\in\R^n$.
    But since the flow map \eqref{subeq:lin_sys_prop_spike_flow} always returns a strictly positive value for the neuron variable $\xi$, there can exist no trajectory $\mathbf{z}^*:[0,\varepsilon]\rightarrow\R^{n+1}$ with $\mathbf{z}^*(0)=(\mathbf{x},\Delta)$ such that $\dot{\mathbf{z}}^*(t)=f(\mathbf{z}^*(t))$ for almost all $t\in[0,\varepsilon]$ and $\mathbf{z}^*(t)\in\mathcal{C}$ for all $t\in(0,\varepsilon]$.
    Indeed, if $\dot{\mathbf{z}}^*(t)=f(\mathbf{z}^*(t))$, then $\xi^*(t)>\Delta$ for any $t\in(0,\varepsilon]$ since $\dot{\xi}^*\ge I_b>0$.
    So point (b) of \cite[Proposition 2.11]{goebel_hybridsystems_2012} is also satisfied and the solution to NICSs from any starting point in $\mathcal{C}\cup\mathcal{D}=\R^n\times\R_{\ge0}$ is unique.
\end{proof}

\begin{proof}[Proof of Lemma~\ref{lemma:jump_emulation}]
    Consider any maximal solution $\mathbf{z}$ of the NICS and any $i\in\N_{\ge1}$.
    For brevity, we write $\dwellt$, $\kappa$, and $\impactmap$ for $\dwellt(\mathbf{x}_i)$, $\kappa(\mathbf{x}_i)$, and $\impactmap(\mathbf{x}_i)$ in this proof.
    The transition matrix of the LTI state can be expressed as
    \begin{equation}
        \impactmap = \impactmap - e^{(A-BKC)\dwellt} + e^{(A-BKC)\dwellt}
    \end{equation}
    which, in norms, gives
    \begin{equation}
        \mnorm{\impactmap} \le \mnorm{e^{(A-BKC)\dwellt}} + \mnorm{\impactmap - e^{(A-BKC)\dwellt}}.
    \end{equation}
    Using the Taylor expansion of both exponentials, the difference between the transition matrix and the exponential can be written as
    \begin{equation}
    \begin{aligned}
        \impactmap-e^{(A-BKC)\dwellt} =&\,\dwellt\Bigl(K-\frac{\alpha}{\dwellt}\Bigr)BC\\&+\sum_{m=2}^{\infty}\frac{\dwellt^m}{m!}\Bigl(A-m\,\frac{\alpha}{\dwellt}BC\Bigr)A^{m-1}\\&-\sum_{m=2}^{\infty}\frac{\dwellt^m}{m!}(A-BKC)^m.
    \end{aligned}
    \end{equation}
    Then, by setting $\kappa=K-\alpha/\dwellt$, developing the first term of each sum together, using the bound $\sum_{m=k}^{\infty}\frac{1}{m!}a^m < a^k\sum_{m=k}^{\infty}\frac{1}{m!}$ valid if $0<a<1$ (with $a=\dwellt\mnorm{A}$ and $a=\dwellt\mnorm{A-BKC}$), and using the identity $\sum_{m=0}^{\infty}\frac{1}{m!}=e$, the constants $C_{1}(\kappa)$, $C_{2}(\kappa)$, and $C_{3}(\kappa)$ of the lemma can be computed.
\end{proof}

\begin{proof}[Proof of Lemma~\ref{lemma:contractive_range}]
    Consider any maximal solution $\mathbf{z}$ of the NICS and any $i\in\N_{\ge1}$, and write $\dwellt$ for $\dwellt(\mathbf{x}_i)$ in this proof.
    From \cite[Section~8.5]{hespanha_linear_2018}, the standard Lyapunov decay estimate gives
    \[
        \mnorm{e^{(A-BKC)\dwellt}}_{P^*}\le e^{-\lambda_0 \dwellt/2},\qquad\lambda_0\eqdefsymbol\frac{\lambda_{\min}(Q^*)}{\lambda_{\max}(P^*)}.
    \]
    Since the constants \eqref{eq:constants} are non-decreasing in $|\kappa|$ and $|\kappa(\mathbf{x}_i)|<\overline{\kappa}$ by \eqref{eq:kappa_bound}, $C_k(\kappa(\mathbf{x}_i))\le\overline{C_k}$ for $k\in\{1,2,3\}$.
    Plugging this in the results of Lemma~\ref{lemma:jump_emulation} expressed in the $P^*$-norm yields
    \begin{equation}
        \mnorm{\impactmap(\mathbf{x}_i)}_{P^*} \le e^{-\lambda_0 \dwellt/2}+\overline{C_1}\dwellt+\overline{C_2}\dwellt^2+\overline{C_3}\dwellt^3,\label{eq:lyapunov_decay_bound}
    \end{equation}
    since $\dwellt<\min\bigl(\mnorm{A}_{P^*}^{-1},\mnorm{A-BKC}_{P^*}^{-1}\bigr)$.

    At $\dwellt=0$ the right-hand side of \eqref{eq:lyapunov_decay_bound} equals one and, since its second derivative with respect to $\dwellt$ is strictly positive for any $\dwellt$, there exists at most one interval $(0, \dwellt^*)$ where \eqref{eq:lyapunov_decay_bound} is smaller than 1, with $\dwellt^* > 0$ the point at which the right-hand side of \eqref{eq:lyapunov_decay_bound} equals one.

    Since the first derivative of the right-hand side of \eqref{eq:lyapunov_decay_bound} at $\dwellt=0$ equals $\overline{C_1}-\lambda_0/2$, the interval only exists when $\overline{C_1}=\overline{\kappa}\mnorm{BC}_{P^*}<\lambda_0/2$, which is the condition on $\overline{\kappa}$ of the lemma.

    Using the inequality $e^{-\lambda_0 \dwellt/2}\le 1-\frac{\lambda_0}{2}\dwellt+\frac{\lambda_0^2}{8}\dwellt^2$, valid for $\dwellt \ge 0$, a lower bound on $\dwellt^*$ can be found by solving the resulting cubic equation. Since $\dwellt=0$ is a root, the cubic factors as $\dwellt\bigl(\overline{C_3}\dwellt^2+(\overline{C_2}+\tfrac{\lambda_0^2}{8})\dwellt+(\overline{C_1}-\tfrac{\lambda_0}{2})\bigr)=0$, and the positive root of the remaining quadratic equation is the bound \eqref{eq:contractive_dwell_time_cond} stated in the lemma.
\end{proof}

\begin{proof}[Proof of Lemma~\ref{lemma:dwell-time}]
    Consider any maximal solution $\mathbf{z}$ of the NICS and any $i\in\N_{\ge1}$, and write $\dwellt$ for $\dwellt(\mathbf{x}_i)$ in this proof.
    Starting from the state $(\mathbf{x}_i,0)$, the LTI state flows as $e^{A\tau}\mathbf{x}_i$, so bounding the dwell time $\dwellt$ amounts to bounding the $\dwellt>0$ that makes the neuron state
    \begin{equation}
        \begin{aligned}
        \int_0^{\dwellt}\dot{\xi}\,\dop{\tau}&=\int_0^{\dwellt}\bigl(|Ce^{A\tau}\mathbf{x}_{i}|+I_b\bigr)\,\dop{\tau}\\&=I_b\dwellt+\int_0^{\dwellt}|Ce^{A\tau}\mathbf{x}_{i}|\,\dop{\tau}
        \end{aligned}\label{eq:bounds_condition}
    \end{equation}
    equal to $\Delta$.

    To compute the upper bound, we can lower bound \eqref{eq:bounds_condition}, which is easily done by ignoring the effect of the input $|C\mathbf{x}|$.

    To compute the lower bound, we can upper bound \eqref{eq:bounds_condition} by using the properties of compatible vector and matrix norms and logarithmic norms, $|Ce^{A\tau}\mathbf{x}_i|\le\mnorm{C}e^{\mu(A)\tau}\vnorm{\mathbf{x}_i}$, then solving the integral. This yields the equation
    \begin{equation}
        I_b\underline{\dwellt}(\mathbf{x}_i)+\frac{\mnorm{C}\vnorm{\mathbf{x}_{i}}}{\mu(A)}\bigl(e^{\mu(A)\underline{\dwellt}(\mathbf{x}_i)}-1\bigr) = \Delta,\label{eq:lower_bound_equality}
    \end{equation}
    assuming $\mu(A)\neq 0$; the case $\mu(A)=0$ reduces to a linear equation.
    This transcendental equation admits a unique real solution, which can be expressed using the principal branch of the Lambert $W$ function.
\end{proof}

\begin{proof}[Proof of Lemma~\ref{lemma:alpha_range_existence}]
    Consider any maximal solution $\mathbf{z}$ of the NICS and any $i\in\N_{\ge1}$.
    A single $\alpha$ with $\alpha/\dwellt(\mathbf{x}_i)\in(a,b)$ for every $\dwellt(\mathbf{x}_i)\in[\underline{\dwellt}(\mathbf{x}_i),\overline{\dwellt}]$ exists if and only if $a\overline{\dwellt}<b\underline{\dwellt}(\mathbf{x}_i)$.
    Using the expressions from Lemma~\ref{lemma:dwell-time} and rearranging, this condition can be expressed as
    \begin{multline}
        a\,\frac{\Delta}{I_b}<b\!\left(\frac{\Delta}{I_b}+\frac{1}{\mu(A)}\left(\frac{\mnorm{C}\vnorm{\mathbf{x}_i}}{I_b}\right.\right.\\\left.\left.-W_0\!\left(\frac{\mnorm{C}\vnorm{\mathbf{x}_i}}{I_b}\,e^{\frac{\mnorm{C}\vnorm{\mathbf{x}_i}}{I_b}}\,e^{\frac{\Delta\mu(A)}{I_b}}\right)\right)\right).
    \end{multline}

    By rearranging and using the upper bound $W_0(x)<\ln(x+1)$, valid for $x>0$, the inequality is implied by
    \begin{equation}
        \frac{\mnorm{C}\vnorm{\mathbf{x}_i}}{I_b} <e^{-(a/b)\frac{\Delta\mu(A)}{I_b}}\!\left(1-e^{(a/b-1)\frac{\Delta\mu(A)}{I_b}}e^{-\frac{\mnorm{C}\vnorm{\mathbf{x}_i}}{I_b}}\right).
    \end{equation}

    Then, using the lower bound $1-e^{-x}>x/(x+1)$, valid for $x>0$, and rearranging, the existence is implied by
    \begin{multline}
        \mnorm{C}\vnorm{\mathbf{x}_i}\,\frac{(1-a/b)\Delta\mu(A)+\mnorm{C}\vnorm{\mathbf{x}_i}}{I_b}-\\e^{-(a/b)\frac{\Delta\mu(A)}{I_b}}\bigl((1-a/b)\Delta\mu(A)+\mnorm{C}\vnorm{\mathbf{x}_i}\bigr)\\<-\mnorm{C}\vnorm{\mathbf{x}_i},
    \end{multline}
    assuming $\mu(A)> 0$.

    Finally, since the inequality is monotone in $I_b$ and is satisfied as $I_b\rightarrow\infty$, the lower bound on $I_b$ satisfying the inequality can be found by solving the equality.
    This transcendental equation admits a unique real solution, expressible using the principal branch of the Lambert $W$ function.
\end{proof}

\begin{proof}[Proof of Lemma~\ref{lemma:x0_decrease_dwell_time}]
    Writing \eqref{eq:lower_bound_equality} for a generic state $\mathbf{x}$ and using implicit differentiation, the derivative of the lower bound from Lemma~\ref{lemma:dwell-time} with respect to $\vnorm{\mathbf{x}}$ can easily be computed as
    \begin{equation}
        \frac{\mathrm{d}\underline{\dwellt}(\mathbf{x})}{\mathrm{d}\vnorm{\mathbf{x}}} =-\frac{\frac{\partial\eqref{eq:lower_bound_equality}}{\partial\vnorm{\mathbf{x}}}}{\frac{\partial\eqref{eq:lower_bound_equality}}{\partial\underline{\dwellt}}}= -\frac{\frac{\mnorm{C}}{\mu(A)}\bigl(e^{\mu(A)\underline{\dwellt}(\mathbf{x})}-1\bigr)}{I_b+\mnorm{C}\vnorm{\mathbf{x}} e^{\mu(A)\underline{\dwellt}(\mathbf{x})}}
    \end{equation}
    and since the numerator and the denominator are strictly positive, the derivative is always strictly negative, assuming $\mu(A)\neq 0$.
\end{proof}
\fi

\section*{Acknowledgment}
\addcontentsline{toc}{section}{Acknowledgment}

Claude Opus 5 and 5.5 (Anthropic) were used to edit the phrasing throughout and to draft initial versions of some passages in Section~\ref{sec:introduction}. 
All technical content, results, and proofs are the authors' own, and the authors reviewed all resulting text and are responsible for it.

\bibliographystyle{IEEEtran}
\bibliography{refs}

\end{document}